\documentclass[preprint,showpacs,aps,superscriptaddress]{revtex4}
\newcommand{\UQ}{ARC Centre of Excellence for Quantum-Atom Optics, 
School of Physical Sciences, University of Queensland, Brisbane, 
QLD 4072, Australia.}

\newcommand{\CPad}{Ecole Normale Sup\'erieure de Physique de Grenoble, Institut National Polytechnique, Grenoble, France.}
 
\usepackage{graphicx,psfrag}
\usepackage{bm}
\newcommand{\etal}{{\em et al.}}
\newcommand{\e}{\mbox{e}}
\newcommand{\pr}{Phys. Rev. }
\newcommand{\jpb}{J. Phys. B }
\newcommand{\opex}{Opt. Express }
\newcommand{\jpa}{J. Phys. A }
\newcommand{\epl}{Europhys. Lett. }

\begin{document}
\title{Tripartite entanglement and threshold properties of coupled intracavity downconversion and sum-frequency generation}

\author{C. Pennarun}
\affiliation{\UQ}
\affiliation{\CPad}
\author{A.~S. Bradley}
\affiliation{\UQ}
\author{M.~K. Olsen}
\affiliation{\UQ}
\date{\today}

\begin{abstract}

The process of cascaded downconversion and sum-frequency generation inside an optical cavity has
been predicted to be a potential source of three-mode continuous-variable entanglement. When the cavity is pumped by two fields, the threshold properties have been analysed, showing that these are more complicated than in well-known processes such as optical parametric oscillation. When there is only a single pumping field, the entanglement properties have been calculated using a linearised fluctuation analysis, but without any consideration of the threshold properties or critical operating points of the system. In this work we extend this analysis to demonstrate that the singly pumped system demonstrates a rich range of threshold behaviour when quantisation of the pump field is taken into account and that asymmetric polychromatic entanglement is available over a wide range of operational parameters.

\end{abstract}

\pacs{42.50.Dv,42.65.Lm,03.65.Ud,03.67.Mn}  

\maketitle

\section{Introduction}
\label{sec:intro}

The modern field of quantum information originally focused on what is known as discrete-variable entanglement and developed to include the study of entanglement between continuous-variable phase quadratures of the electromagnetic field, which have a close analogy with the original position and momentum considered in the famous Einstein-Podolsky-Rosen (EPR) paradox~\cite{EPR}. In the beginning this research considered bipartite entanglement as produced by, for example, the optical parametric oscillator (OPO)~\cite{eprMDR} and led to experimental demonstrations of the EPR paradox~\cite{Pereira} and of what is known as continuous-variable quantum teleportation~\cite{teleport1,teleport2,teleport3}. Many systems have now been studied, both theoretically and experimentally, with continuous-variable bipartite entanglement now considered an important resource for quantum information applications~\cite{Braunstein}.

Recently there has been much attention paid to the production of continuous-variable tripartite
entanglement, obtained either by mixing squeezed beams on unbalanced beamsplitters~\cite{Jing,Aoki}, or via the interaction of multiple input beams in nonlinear media with cascaded or concurrent $\chi^{(2)}$ nonlinearities. Among the latter are systems using either single~\cite{nussen}, twin~\cite{Ferraro,Guo,Olsen,Bondani,Olsen2} or triple~\cite{Pooser,Pfister,EPR3} nonlinearities. These nonlinear processes have been analysed and demonstrated in both the travelling-wave and intracavity configurations. In this work we are interested in an intracavity process which combines parametric downconversion with sum frequency generation, as theoretically analysed by Yu \etal~\cite{Yu}. The idea of combining these two processes is due to Smithers and Lu~\cite{Smithers}, who did not consider enclosing the processes in an optical cavity. The intracavity process with two pump fields was first analysed by Guo \etal~\cite{Guo}, who used quantum Langevin equations~\cite{Langevin} with an undepleted pump approximation, which can give no insight into threshold behaviour or any critical operating points. An analysis which included quantisation of the two pump fields was performed by Olsen and Bradley~\cite{Olsen}, demonstrating that the system had quite different stability and threshold behaviours to the normal OPO. As Yu \etal~\cite{Yu} have also used an undepleted pump approximation (also known as the parametric approximation) and quantum Langevin equations for the singly pumped intracavity system, they are not able to determine the threshold behaviours or the stability of the equations they use. In this paper we apply a fully quantised treatment of all the interacting fields, finding that there are two separate parameter regimes, one of which has an oscillation threshold while the other does not. The reason for this behaviour, which is more complicated than that of the standard OPO, is that downconversion considered separately does exhibit a threshold while sum frequency conversion does not. As we will demonstrate below, it is the competition and interplay of these two processes which leads to more complicated behaviour.

\section{Hamiltonian and equations of motion}
\label{sec:Ham}

Our system is a nonlinear medium inside an optical cavity which is pumped at frequency $\omega_{0}$ and is resonant at all the frequencies involved. In Yu \etal~\cite{Yu} the nonlinear medium is a quasiperiodic superlattice. In the downconversion part of the intracavity process, two fields at  
$\omega_{1}$ and $\omega_{3}$ are generated, where $\omega_{0}=\omega_{1}+\omega_{3}$. We will denote the effective nonlinearity for this process by $\chi_{1}$. The pump field at $\omega_{0}$ can then combine with the field at $\omega_{3}$ in a sum frequency generation process, to produce a further field at $\omega_{2}$, with the effective nonlinearity represented as $\chi_{2}$. We will use the annihilation operator $\hat{b}$ to describe the field at $\omega_{0}$, while the operators $\hat{a}_{j}$ will be used for the fields at $\omega_{j}$.

The Hamiltonian can be written as
\begin{equation}
\hat{H}_{\rm tot}=\hat{H}_{\rm pump}+\hat{H}_{\rm int}+\hat{H}_{\rm damp},
\label{eq:H}
\end{equation}
where the interaction Hamiltonian is
\begin{equation}\label{eq:Hint}
\hat{H}_{\rm int} =
i\hbar\chi_{1}(\hat{b}\hat{a}_1^\dag\hat{a}_3^\dag-\hat{b}^\dag\hat{a}_1\hat{a}_3)+i\hbar\chi_{2}(\hat{b}\hat{a}_3\hat{a}_2^\dag-\hat{b}^\dag\hat{a}_3^\dag\hat{a}_2),
\label{eq:ham}
\end{equation}
the Hamiltonian describing the cavity pumping is 
\begin{equation}
\label{eq:Hpump}
\hat{H}_{\rm pump} =
i\hbar(\epsilon\hat{b}^\dag-\epsilon^{*}\hat{b}),
\end{equation}
and the cavity damping Hamiltonian is
\begin{equation}
\label{eq:Hdamp}
\hat{H}_{\rm damp} =
\hbar(\hat{\Gamma}_0\hat{b}^\dag+\hat{\Gamma}_0^\dag\hat{b})+\hbar\sum_{j=1:3}(\hat{\Gamma}_j\hat{a}_j^\dag+\hat{\Gamma}_j^\dag\hat{a}_j).
\end{equation}
In the above, $\epsilon$ is the pump field which enters the cavity, which will be described classically, and the $\Gamma_j$ are reservoir operators for each of the intracavity modes.

To calculate the fluctuation and entanglement properties of the system we will derive fully quantum equations of motion using the positive-P pseudprobability distribution~\cite{P+}, as this naturally allows us to calculate the normally-ordered operator expectation values required to find output spectra. Proceeding via the normal methods~\cite{Crispin} and making the zero temperature and Markov approximations for the reservoir~\cite{Danbook}, we find the Fokker-Planck equation for the P function~\cite{Roy,Sud} of the system as
\begin{eqnarray}
\frac{dP}{dt} &=& \left\{-\left[\frac{\partial}{\partial\alpha_{1}}\left(\gamma_{1}\alpha_{1}-\chi_{1}\alpha_{3}^{\ast}
\beta\right) + \frac{\partial}{\partial\alpha_{1}^{\ast}}\left(\gamma_{1}\alpha_{1}^{\ast}-\chi_{1}\alpha_{3}\beta^{\ast}
\right)\right.\right.\nonumber\\
& & 
\left.\left.
+\frac{\partial}{\partial\alpha_{2}}\left(\gamma_{2}\alpha_{2}-\chi_{2}\alpha_{3}\beta\right) + 
\frac{\partial}{\partial\alpha_{2}^{\ast}}\left(\gamma_{2}\alpha_{2}^{\ast}-
\chi_{2}\alpha_{3}^{\ast}\beta^{\ast}\right)\right.\right.\nonumber\\
& &
\left.\left.
+\frac{\partial}{\partial\alpha_{3}}\left(\gamma_{3}\alpha_{3}-\chi_{1}\alpha_{1}^{\ast}\beta+
\chi_{2}\alpha_{2}\beta^{\ast}\right)
+\frac{\partial}{\partial\alpha_{3}^{\ast}}\left(\gamma_{3}\alpha_{3}^{\ast}-\chi_{1}\alpha_{1}\beta^{\ast}+
\chi_{2}\alpha_{2}^{\ast}\beta\right)\right.\right.\nonumber\\
& &
\left.\left.
+\frac{\partial}{\partial\beta}\left(\gamma_{0}\beta-\epsilon+\chi_{1}\alpha_{1}\alpha_{3}+\chi_{2}\alpha_{2}\alpha_{3}^{\ast}\right)\right.\right.\nonumber\\
& &
\left.\left.
+\frac{\partial}{\partial\beta^{\ast}}\left(\gamma_{0}\beta^{\ast}-\epsilon^{\ast}+\chi_{1}\alpha_{1}^{\ast}\alpha_{3}^{\ast}+\chi_{2}\alpha_{2}^{\ast}\alpha_{3}\right)\right]\right.\nonumber\\
& & 
\left.
+\frac{1}{2}\left[2\chi_{1}\left(\frac{\partial^{2}}{\partial\alpha_{1}\partial\alpha_{3}}\beta+\frac{\partial^{2}}{\partial\alpha_{1}^{\ast}\partial\alpha_{3}^{\ast}}\beta^{\ast}\right)-2\chi_{2}\left(\frac{\partial^{2}}{\partial\alpha_{3}\partial\beta}\alpha_{2}^{\ast}+\frac{\partial^{2}}{\partial\alpha_{3}^{\ast}\partial\beta^{\ast}}\alpha_{2}\right)\right]
\right\}P,
\label{eq:fokkerplanck}
\end{eqnarray}
where the $\gamma_{j}$ are the cavity loss rates at frequency $\omega_{j}$.
As this Fokker-Planck equation does not possess a positive-definite diffusion matrix, we must double the phase-space and use the positive-P representation to find the appropriate stochastic differential equations. This results in replacement of the conjugate variables by $\alpha_{j}^{+}$ and $\beta^{+}$, which are the complex conjugates of the uncrossed variables only in the mean. Stochastic averages of products of these variables are then equal to normally-ordered expectation values of the corresponding operators. We find the coupled set of stochastic differential equations as
\begin{eqnarray}
\frac{d\alpha_{1}}{dt} &=& -\gamma_{1}\alpha_{1}+\chi_{1}\alpha_{3}^{+}\beta+\sqrt{\frac{\chi_{1}\beta}{2}}(\eta_{1}+i\eta_{2}),\nonumber\\
\frac{d\alpha_{1}^{+}}{dt} &=& -\gamma_{1}\alpha_{1}^{+}+\chi_{1}\alpha_{3}\beta^{+}+\sqrt{\frac{\chi_{1}\beta^{+}}{2}}(\eta_{3}+i\eta_{4}),
\nonumber\\
\frac{d\alpha_{2}}{dt} &=& - \gamma_{2}\alpha_{2}+\chi_{2}\alpha_{3}\beta,\nonumber\\
\frac{d\alpha_{2}^{+}}{dt} &=& - \gamma_{2}\alpha_{2}^{+}+\chi_{2}\alpha_{3}^{+}\beta^{+},\nonumber\\
\frac{d\alpha_{3}}{dt} &=& -\gamma_{3}\alpha_{3}+\chi_{1}\alpha_{1}^{+}\beta-\chi_{2}\alpha_{2}\beta^{+}+\sqrt{\frac{\chi_{1}\beta}{2}}(\eta_{1}-i\eta_{2})+\sqrt{-\frac{\chi_{2}\alpha_{2}^{+}}{2}}(\eta_{5}+i\eta_{6}),\nonumber\\
\frac{d\alpha_{3}^{+}}{dt} &=& -\gamma_{3}\alpha_{3}^{+}+\chi_{1}\alpha_{1}\beta^{+}-\chi_{2}\alpha_{2}^{+}\beta+\sqrt{\frac{\chi_{1}\beta^{+}}{2}}(\eta_{3}-i\eta_{4})+\sqrt{-\frac{\chi_{2}\alpha_{2}}{2}}(\eta_{7}+i\eta_{8}),\nonumber\\
\frac{d\beta}{dt} &=& \epsilon - \gamma_{0}\beta-\chi_{1}\alpha_{1}\alpha_{3}-\chi_{2}\alpha_{2}\alpha_{3}^{+}+\sqrt{\frac{-\chi_{2}\alpha_{2}^{+}}{2}}(\eta_{5}-i\eta_{6}),\nonumber\\
\frac{d\beta^{+}}{dt} &=& \epsilon^{\ast} - \gamma_{0}\beta^{+}-\chi_{1}\alpha_{1}^{+}\alpha_{3}^{+}-\chi_{2}\alpha_{2}^{+}\alpha_{3}+\sqrt{\frac{-\chi_{2}\alpha_{2}}{2}}(\eta_{7}-i\eta_{8}),
\label{eq:SDE}
\end{eqnarray}
where the $\eta_{j}$ are real Gaussian noise terms with the properties
\begin{equation}
\overline{\eta_{j}(t)}=0,\:\:\:\overline{\eta{j}(t)\eta_{k}(t')}=\delta_{jk}\delta(t-t').
\label{eq:ruido}
\end{equation}

In cases where the procedure is valid, the noise terms may be dropped and the resulting semiclassical equations linearised about their steady states, which results in the process being treated as an Ornstein-Uhlenbeck process~\cite{SMCrispin}, allowing for easy calculation of the output spectra. The validity of this linearised fluctuation analysis is usually found by calculating the eigenvalues of the resulting drift matrix for the fluctuations and requires knowledge of the classical steady-state solutions. In fact, in the present case, we find that stochastic integration of the above equations presents various stability problems in the regions where they cannot be linearised, so that in section~\ref{sec:Wigner} we will turn to the truncated Wigner representation~\cite{Robert} to find time domain solutions in these parameter regimes.

\section{Linearised fluctuation analysis}
\label{sec:linfluctuations}

In the steady-state, we can always decompose the system variables into their mean values and a part which fluctuates about these. In many cases the mean value solutions of the noiseless equations are equal to the operator expectation values and the fluctuations can be treated as being stable and Gaussian about zero means. In these cases we may use this linearised fluctuation analysis as a simple method to calculate measurable spectra. We will perform this process on the positive-P equations, beginning with the decomposition $\alpha_{i}=\overline{\alpha_{i}}+\delta\alpha_{i}$ and similarly for $\beta$. This gives us the set of equations for the fluctuating terms
\begin{eqnarray}
\frac{d}{dt}\delta\alpha_{1} &=& -\gamma_{1}\delta\alpha_{1}+\chi_{1}\overline{\beta}\delta\alpha_{3}^{\ast}+\sqrt{\frac{\chi_{1}\overline{\beta}}{2}}(\eta_{1}+i\eta_{2}),\nonumber\\
\frac{d}{dt}\delta\alpha_{1}^{\ast} &=& -\gamma_{1}\delta\alpha_{1}^{\ast}+\chi_{1}\overline{\beta^{\ast}}\delta\alpha_{3}+\sqrt{\frac{\chi_{1}\overline{\beta^{\ast}}}{2}}(\eta_{3}+i\eta_{4}),\nonumber\\
\frac{d}{dt}\delta\alpha_{2} &=& -\gamma_{2}\delta\alpha_{2}+\chi_{2}\overline{\beta}\delta\alpha_{3},\nonumber\\
\frac{d}{dt}\delta\alpha_{2}^{\ast} &=& -\gamma_{2}\delta\alpha_{2}^{\ast}+\chi_{2}\overline{\beta^{\ast}}\delta\alpha_{3}^{\ast},\nonumber\\
\frac{d}{dt}\delta\alpha_{3} &=& -\gamma_{3}\delta\alpha_{3}+\chi_{1}\overline{\beta}\delta\alpha_{1}^{\ast}-\chi_{2}\overline{\beta^{\ast}}\delta\alpha_{2}+\sqrt{\frac{\chi_{1}\overline{\beta}}{2}}(\eta_{1}-i\eta_{2}),\nonumber\\
\frac{d}{dt}\delta\alpha_{3}^{\ast} &=& -\gamma_{3}\delta\alpha_{3}^{\ast}+\chi_{1}\overline{\beta^{\ast}}\delta\alpha_{1}-\chi_{2}\overline{\beta}\delta\alpha_{2}^{\ast}+\sqrt{\frac{\chi_{1}\overline{\beta^{\ast}}}{2}}(\eta_{3}-i\eta_{4}),\nonumber\\
\frac{d}{dt}\delta\beta &=& -\gamma_{0}\delta\beta,\nonumber\\
\frac{d}{dt}\delta\beta^{\ast} &=& -\gamma_{0}\delta\beta^{\ast},
\label{eq:linfluk}
\end{eqnarray}
which may be written in matrix form for the vector
\begin{equation}
\delta\tilde{\alpha}=\left[\delta\alpha_{1},\delta\alpha_{1}^{\ast},\delta\alpha_{2},\delta\alpha_{2}^{\ast},\delta\alpha_{3},\delta\alpha_{3}^{\ast},\delta\beta,\delta\beta^{\ast}\right]^{T},
\label{eq:ashvek}
\end{equation}
as
\begin{equation}
\frac{d}{dt}\delta\tilde{\alpha} = -A\,\delta\tilde{\alpha}+B\,d\tilde{W},
\label{eq:ornstein}
\end{equation}
where $A$ is the drift matrix, $B$ contains the steady-state coefficients of the noise terms,  and $d\tilde{W}$ is a vector of Wiener increments. The condition for stability of the fluctuations is that the eigenvalues of $A$ have no negative real parts. When this condition is fulfilled, we may calculate the intracavity spectral matrix as
\begin{equation}
S(\omega) = \left(A+i\omega\openone\right)^{-1}BB^{T}\left(A^{T}-iw\openone\right)^{-1},
\label{eq:Sin}
\end{equation}
which, along with the well-known input-output relations~\cite{Langevin}, allow us to calculate the measurable spectral quantities outside the cavity.

\subsection{Steady-state classical solutions}
\label{subsec:stationary}

The classical equations for the mean values are found as
\begin{eqnarray}
\frac{d\overline{\alpha}_1}{dt} &=& -\gamma_1\overline{\alpha}_1 + \chi_1\overline{\alpha}_3^{\ast}\overline{\beta},\nonumber\\
\frac{d\overline{\alpha}_2}{dt} &=& -\gamma_2\overline{\alpha}_2 + \chi_2\overline{\alpha}_3\overline{\beta},\nonumber\\
\frac{d\overline{\alpha}_3}{dt} &=& -\gamma_3\overline{\alpha}_3 + \chi_1\overline{\alpha}_1^{\ast}\overline{\beta} -
\chi_2\overline{\alpha}_2\overline{\beta}^{\ast},\nonumber\\
\frac{d \overline{\beta}}{dt} &=& \epsilon - \gamma_0\overline{\beta} - \chi_1\overline{\alpha}_1\overline{\alpha}_3 -
\chi_2\overline{\alpha}_2\overline{\alpha}_3^{\ast},
\label{eq:steady}
\end{eqnarray}
and may be solved for the steady-state solutions which enable us to perform the necessary stability analysis. We find that the solutions are divided into two different classes, depending on whether an oscillation threshold is present or not.

\subsubsection{Regime with threshold}
\label{susubsec:limiar}

If $\chi_1^2\gamma_2 > \chi_2^2\gamma_1$, we find that the system has a threshold pumping value below which it will not oscillate. If the value of the pump field $\epsilon$
is below
\begin{equation}
\epsilon_c = \frac{\gamma_0\sqrt{\gamma_3}}{\sqrt{\frac{\chi_1^2}{\gamma_1} - \frac{\chi_2^2}{\gamma_2}}},
\label{eq:bombeio}
\end{equation}
the signal modes will not be macroscopically occupied. We note that this is totally different from the expression which would be expected if we considered the threshold for the downconversion process in isolation, which would be given as $\epsilon_{c}^{o}=\gamma_{0}\sqrt{\gamma_{1}\gamma_{3}}/\chi_{1}$. This difference in threshold cannot be calculated in the approach taken by Yu \etal~\cite{Yu}. 
The analytical expressions for the different  mean-value
steady-state solutions are (note we will now drop the bar over the variables for notational convenience) :

(i) $\epsilon < \epsilon_c$
\begin{eqnarray}
\beta^{ss} &=& \frac{\epsilon}{\gamma_0}, \nonumber \\
\alpha_{j}^{ss} &=& 0,
\label{eq:belowthresh}
\end{eqnarray}
where $j=1,2,3$.

(ii) $\epsilon > \epsilon_c$
\begin{eqnarray}
\beta^{ss} &=& \sqrt{\frac{\gamma_3}{\frac{\chi_1^2}{\gamma_1} - \frac{\chi_2^2}{\gamma_2}}}, \nonumber \\
\alpha_{1}^{ss} &=& \pm\frac{\chi_1}{\gamma_1}\beta^{ss}\sqrt{\frac{\epsilon - \epsilon_c}{\frac{\epsilon_c}{\gamma_0}(\frac{\chi_1^2}{\gamma_1} +
\frac{\chi_2^2}{\gamma_2})}}\e^{-i\theta} , \nonumber \\
\alpha_{2}^{ss} &=& \pm\frac{\chi_2}{\gamma_2}\beta^{ss}\sqrt{\frac{\epsilon - \epsilon_c}{\frac{\epsilon_c}{\gamma_0}(\frac{\chi_1^2}{\gamma_1} +
\frac{\chi_2^2}{\gamma_2})}}\e^{i\theta} , \nonumber \\
\alpha_{3}^{ss} &=& \pm\sqrt{\frac{\epsilon - \epsilon_c}{\frac{\epsilon_c}{\gamma_0}(\frac{\chi_1^2}{\gamma_1} +
\frac{\chi_2^2}{\gamma_2})}}\e^{i\theta} ,
\label{eq:abovethresh}
\end{eqnarray}
where $\theta$ is an undetermined phase.
We notice that due to the presence of the square root, the sign of these solutions is unknown. However, inspection shows that the square roots all have to be the same sign, whether this is positive or negative. Nevertheless, because the only phase we know is the phase of the pump field $\epsilon$ which
we take as real, $\beta^{ss}$ will also be real. As $\theta$ is not fixed, these above threshold solutions will exhibit phase diffusion, as previously found in the nondegenerate parametric oscillator~\cite{reid}.

\subsubsection{Regime without threshold}
\label{subsubsec:semlimiar}

We find that if $\chi_1^2\gamma_{2} < \chi_2^2\gamma_{1}$ there is no threshold predicted by the classical equations, which means that for any value of the
pump field the signal modes will not be macroscopically occupied. The expressions for the steady-state solutions are the
same as the below threshold solutions of the previous case,
\begin{eqnarray}
\beta^{ss} &=& \frac{\epsilon}{\gamma_0}, \nonumber \\
\alpha_{j}^{ss} &=& 0.
\label{eq:belowthresh2}
\end{eqnarray}
In our analyses in this regime we will scale the pump amplitude by the normal OPO threshold, $\epsilon^{o}_c =
\gamma_0 \sqrt{\gamma_1 \gamma_{3}} / \chi_1$, so that the pump field
$\epsilon$ will be expressed as a proportion of this threshold.
Now that the classical steady-state values in the different areas are known, we can analyse the stability of the
fluctuations.

\subsection{Stability analysis}
\label{subsec:stabilidade}

To determine the validity of the linearisation process we will now analyse the eigenvalues of the drift matrix $A$ of Eq.~\ref{eq:ornstein}. This is written out in full as
\begin{equation}
A=\left[\begin{array}{cccccccc}
\gamma_{1} & 0 & 0 & 0 & 0 & -\chi_1\beta^{ss} & -\chi_1\alpha_{3}^{\ast} & 0 \\
0 & \gamma_1 & 0 & 0 & -\chi_1(\beta^{ss})^{\ast} & 0 & 0 & -\chi_1\alpha_3 \\
0 & 0 & \gamma_2 & 0 & -\chi_2\beta^{ss} & 0 & -\chi_2\alpha_{3}^{ss} & 0 \\
0 & 0 & 0 & \gamma_2 & 0 & -\chi_2(\beta^{ss})^{\ast} & 0 & -\chi_2(\alpha_{3}^{ss})^{\ast} \\
0 & -\chi_1\beta^{ss} & \chi_2(\beta^{ss})^{\ast} & 0 & \gamma_3 & 0 & -\chi_1(\alpha_{1}^{ss})^{\ast} &
\chi_2\alpha_{2}^{ss} \\
-\chi_1(\beta^{ss})^{\ast} & 0 & 0 & \chi_2\beta^{ss} & 0 & \gamma_3 & \chi_2(\alpha_{2}^{ss})^{\ast} &
-\chi_1\alpha_{1}^{ss} \\
\chi_1\alpha_{3}^{ss} & 0 & \chi_2(\alpha_{3}^{ss})^{\ast} & 0 & \chi_1\alpha_{1}^{ss} & \chi_2\alpha_{2}^{ss} & \gamma_0 & 0 \\
0 & \chi_1(\alpha_{3}^{ss})^{\ast} & 0 & \chi_2\alpha_{3}^{ss} & \chi_2(\alpha_{2}^{ss})^{\ast} & \chi_1(\alpha_{1}^{ss})^{\ast} & 0 & \gamma_0
\end{array}\right].
\label{eq:ssAmat}
\end{equation}

The analytical expressions for the eigenvalues of the matrix $A$ above threshold are not easily obtained, but below threshold
where all the $\alpha_{j}^{ss}$ are zero and $\beta^{ss}=\epsilon/\gamma_{0}$, we find the characteristic polynomial as
\begin{equation}
(\gamma_0 - \lambda)^2\left[(\gamma_1 - \lambda)(\gamma_2 - \lambda)(\gamma_3 - \lambda) +
\lambda\beta^{ss\,2}(\chi_1^2 - \chi_2^2) + \beta^{ss\,2}(\gamma_1\chi_2^2 - \gamma_2\chi_1^2)\right]^2 = 0,
\label{eq:characteristic}
\end{equation}
By studying the variation of this function we note that the system is always stable below threshold (whether $\chi_1^2\gamma_2 > \chi_2^2\gamma_1$
or not) and unstable at threshold (for $\beta^{ss} = \epsilon_c /
\gamma_0$). In the special case that $\gamma_{1}=\gamma_{2}=\gamma_{3}=\gamma$ we may find simple analytical solutions as
\begin{eqnarray}
\lambda_{1,2} &=& \gamma_{0},\nonumber\\
\lambda_{3,4} &=& \gamma, \nonumber\\
\lambda_{5,6} &=& \gamma + \frac{\epsilon}{\gamma_{0}}\sqrt{\chi_{1}^{2}-\chi_{2}^{2}},\nonumber\\
\lambda_{7,8} &=& \gamma - \frac{\epsilon}{\gamma_{0}}\sqrt{\chi_{1}^{2}-\chi_{2}^{2}}.
\label{eq:autovalores}
\end{eqnarray}
It is immediately obvious that only $\lambda_{7,8}$ can possibly have a negative real part in this special case, which will happen when
\begin{equation}
\epsilon^{2}>\frac{\gamma_{0}^{2}\gamma^{2}}{\chi_{1}^{2}-\chi_{2}^{2}},
\label{eq:specialcritical}
\end{equation}
and means that any fluctuations will tend to grow, invalidating any linearised fluctuation analysis in this regime.
We see that this is consistent with the critical pump value given above, in Eq.~\ref{eq:bombeio}.
We find numerically that the fluctuations above threshold cannot be linearised due to the presence of a zero eigenvalue.

\begin{figure}[tbhp]
\includegraphics[width=.75\columnwidth]{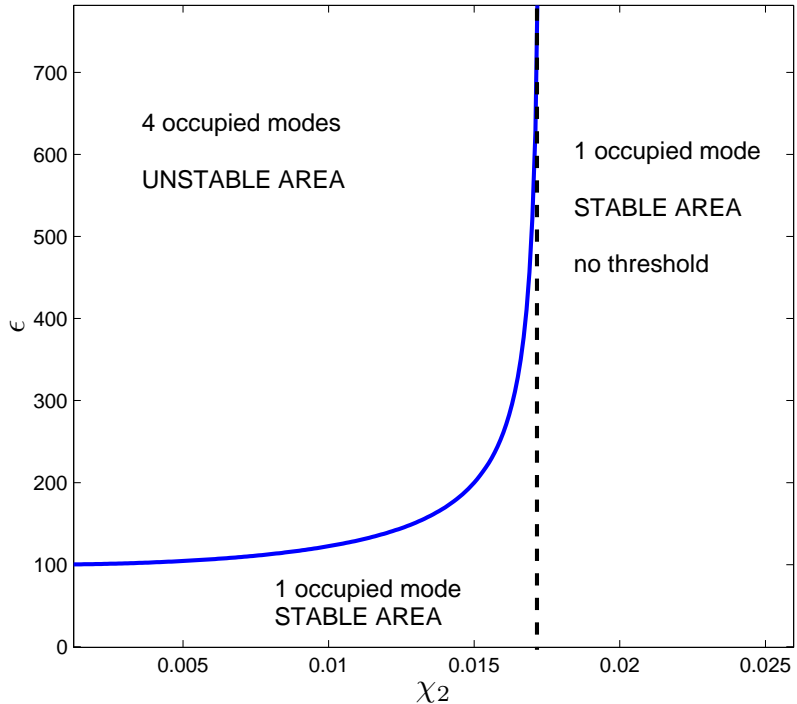}
\caption{(colour online) Stability of the steady state solutions with $\chi_1=0.01$,
$\gamma_0 = \gamma_1 = \gamma_3 = 1$, and $\gamma_2 = 3$, as $\chi_{2}$ and the pump amplitude are varied. The dashed line
shows the separation between the system with and without threshold.}
\label{fig:number}
\end{figure}

In Fig.~\ref{fig:number}, we give a plot of the different stability regions, for $\gamma_0 = \gamma_1 = \gamma_3 = 1$, $\gamma_2 = 3\gamma_1$, and $\chi_1 = 0.01\gamma_1$, as $\chi_2$ and $\epsilon$ are varied. 
We see that when the system is oscillating and the non-pump modes are occupied, it is always unstable and must be treated numerically using stochastic equations. In the stable region below and to the right of the solid line we may use a linearised fluctuation analysis to calculate the correlations of interest. 

\section{Detection of tripartite entanglement}
\label{sec:tripart}

There are a number of inequalities whose violation is sufficient to demonstrate the existence of continuous-variable tripartite entanglement, all of which are based on the inseparability of the system density matrix. Unlike bipartite entanglement, where two modes are either entangled or not, there are a number of cases to be considered, depending on possible partitions of the density matrix~\cite{Giedke}. In this work we are interested in the case where the density matrix is not separable in any form, often known as genuine tripartite entanglement. Before we describe
the criteria we will use here, we need to define the quadrature operators we will use, as different normalisations exist in the literature and can alter the exact form of the inequalities used. As we are considering that the cavity will be at resonance for all modes, we may use the orthogonal quadrature definitions,
\begin{eqnarray}
\hat{X}_{j} &=& \hat{a}_j + \hat{a}_{j}^{\dag},\nonumber\\
\hat{Y}_{j} &=& -i(\hat{a}_{j} - \hat{a}_{j}^{\dag}),
\label{eq:operator}
\end{eqnarray}
with the Heisenberg uncertainty principle requiring that $V(\hat{X}_{i})V(\hat{Y}_{i}) \geq 1$. We note here that any cavity detuning or Kerr interaction can change the quadrature angle at which the best quantum correlations are found~\cite{granja,nlc}, but this is not generally the case for a resonant cavity with $\chi^{(2)}$ interactions. 

We will use two different sets of conditions to investigate the presence of entanglement in this system, both of which were described by van Loock and Furusawa~\cite{vanLoock}. The first of these gives a set of inequalities,
\begin{eqnarray}
V_{12} &=& V(\hat{X}_{1} - \hat{X}_{2}) + V(\hat{Y}_1 +\hat{Y}_2 + g_3\hat{Y}_3) \geq 4,\nonumber\\
V_{13} &=& V(\hat{X}_{1} - \hat{X}_{3}) + V(\hat{Y}_1 + g_2\hat{Y}_2 + \hat{Y}_3) \geq 4,\nonumber\\
V_{23} &=& V(\hat{X}_{2} - \hat{X}_{3}) + V(g_1\hat{Y}_1 + \hat{Y}_2 + \hat{Y}_3) \geq 4,
\label{eq:VLFtripart}
\end{eqnarray}
the violation of any two of which shows that the system is fully inseparable and genuine
tripartite entanglement in guaranteed.
The $g_{i}$ are arbitrary real numbers which may be chosen to minimise the correlations, and will be optimised here as was done in Ref.~\cite{Olsen2}, giving
\begin{eqnarray}
g_{1} &=& -\frac{V(\hat{Y}_{1},\hat{Y}_{2})+V(\hat{Y}_{1},\hat{Y}_{3})}{V(\hat{Y}_{1})},\nonumber\\
g_{2} &=& -\frac{V(\hat{Y}_{1},\hat{Y}_{2})+V(\hat{Y}_{2},\hat{Y}_{3})}{V(\hat{Y}_{2})},\nonumber\\
g_{3} &=& -\frac{V(\hat{Y}_{1},\hat{Y}_{3})+V(\hat{Y}_{2},\hat{Y}_{3})}{V(\hat{Y}_{3})}.
\label{eq:optimumgees}
\end{eqnarray}

The second conditions provide inequalities for which, if any one is violated, genuine tripartite entanglement is demonstrated. They are
\begin{eqnarray}
V_{123} &=& V(\hat{X}_{1} - \frac{\hat{X}_{2} + \hat{X}_{3}}{\sqrt{2}}) + V(\hat{Y}_1 + \frac{\hat{Y}_2 +
\hat{Y}_3}{\sqrt{2}}) \geq 4,\nonumber\\
V_{312} &=& V(\hat{X}_{3} - \frac{\hat{X}_{1} + \hat{X}_{2}}{\sqrt{2}}) + V(\hat{Y}_3 + \frac{\hat{Y}_1 +
\hat{Y}_2}{\sqrt{2}}) \geq 4,\nonumber\\
V_{231} &=& V(\hat{X}_{2} - \frac{\hat{X}_{3} + \hat{X}_{1}}{\sqrt{2}}) + V(\hat{Y}_2 + \frac{\hat{Y}_3 +
\hat{Y}_1}{\sqrt{2}}) \geq 4.
\label{eq:tripart2}
\end{eqnarray}
As in previous cases where the system is described by an asymmetric Hamiltonian (i.e. mode indices cannot be swapped without changing the system), the correct choice of indices during the measurement is important for both sets of correlations given above~\cite{Olsen2}. 

All these correlations can be simply calculated from the intracavity spectral matrix of Eq.~\ref{eq:Sin} and the use of the standard input-output relations~\cite{Langevin} to give the measurable spectra outside the cavity. For example, spectral variances and covariances are calculated as
\begin{eqnarray}
S_{X_{j}}^{out}(\omega) &=& 1+2\gamma_{j}S_{X_{j}}(\omega),\nonumber\\
S_{X_{j},X_{k}}^{out}(\omega) &=& 2\sqrt{\gamma_{j}\gamma_{k}}S_{X_{j},X_{k}}(\omega),
\label{eq:Sinout}
\end{eqnarray}
and similarly for the $\hat{Y}$ quadratures. As this notation can become rather clumsy, we will use $S_{ij}$ and $S_{ijk}$ in what follows to refer to the output spectral qualities equivalent to the $V_{ij}$ and $V_{ijk}$ correlations defined above. The same inequalities hold for these.

\section{Spectral results in the stable regime}
\label{sec:stablespek}

\begin{figure}[tbhp]
\includegraphics[width=0.75\columnwidth]{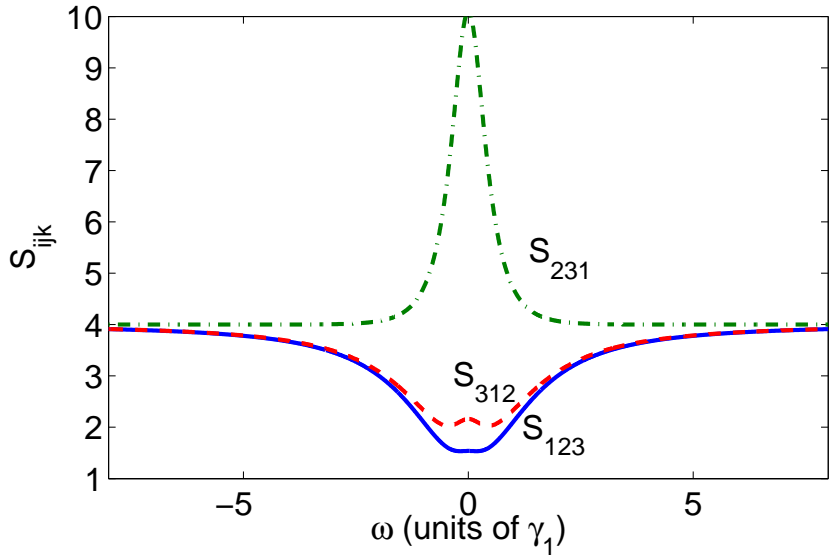}
\caption{(colour online) $S_{ijk}$ criteria below threshold for $\epsilon = 0.5\epsilon_c$, $\gamma_0 =\gamma_{1} = \gamma_{3} = 1$, and $\gamma_{2} = 3$. In this and all subsequent graphs, the results are dimensionless.}
\label{fig:halfcrit1}
\end{figure}

\begin{figure}
\includegraphics[width=.75\columnwidth]{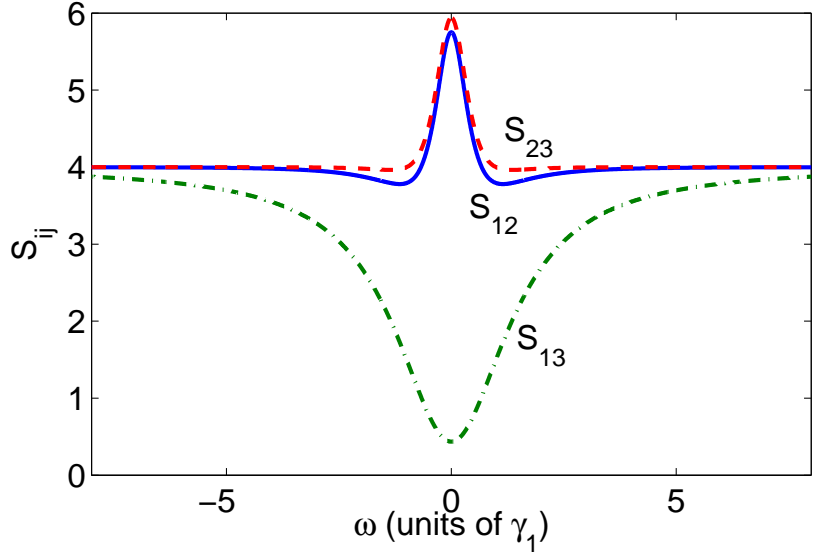}
\caption{(colour online) $S_{ij}$ criteria below threshold for $\epsilon = 0.5\epsilon_c$, $\gamma_0 =\gamma_{1} = \gamma_{3} = 1$, and $\gamma_{2} = 3$.}
\label{fig:halfcrit2}
\end{figure}

Although it is possible to obtain analytical results for the $S_{ij}$ and $S_{ijk}$, these are extremely unwieldy and not at all enlightening. We have therefore chosen to present the results graphically for various parameter regimes.
In Fig.~\ref{fig:halfcrit1} and Fig.~\ref{fig:halfcrit2}, we show the results for the two different types of correlations at half the critical pumping amplitude, in the regime where $\chi_{2}=0.4\chi_{1}$ and with the loss rates set as $\gamma_0 =\gamma_{1} = \gamma_{3} = 1$, and $\gamma_{2} = 3$. In all results presented here we have used a value of $\chi_{1}=0.01$. In Fig.~\ref{fig:halfcrit1} we see clear evidence of genuine tripartite entanglement, with both $S_{123}$ and $S_{312}$ obviously violating the inequality, with only one of these being below $4$ already being sufficient. In Fig.~\ref{fig:halfcrit2}, where two of the inequalities need to be violated, we see that $S_{12}$ and $S_{13}$ both show entanglement, although not over as large a frequency region as the $S_{ijk}$. As in Ref.~\cite{Olsen2}, this is a result of the asymmetry of the Hamiltonian and the fact that a violation of the tripartite inequalities is a sufficient but not necessary condition for the demonstration of tripartite entanglement.
 
\begin{figure}[tbhp]
\includegraphics[width=0.75\columnwidth]{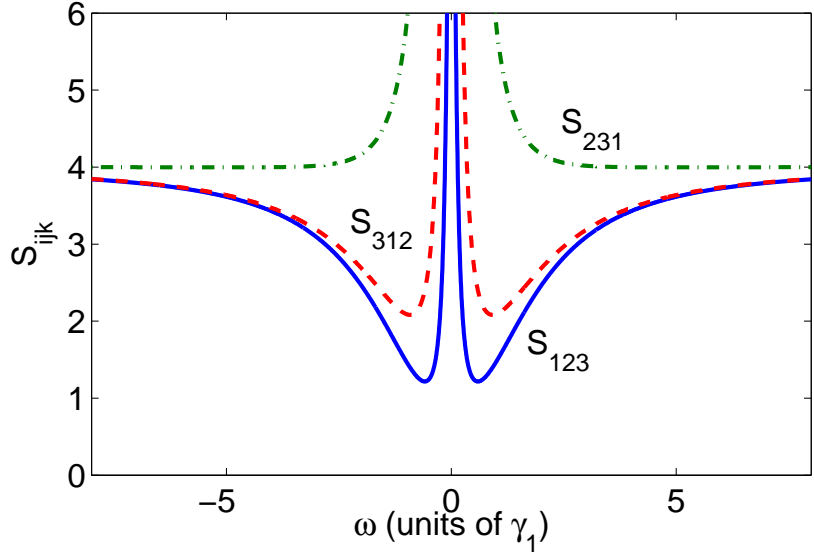}
\caption{(colour online) $S_{ijk}$ criteria below threshold with $\epsilon = 0.9\epsilon_c$.}
\label{fig:experiment1}
\end{figure}

\begin{figure}[tbhp]
\includegraphics[width=0.75\columnwidth]{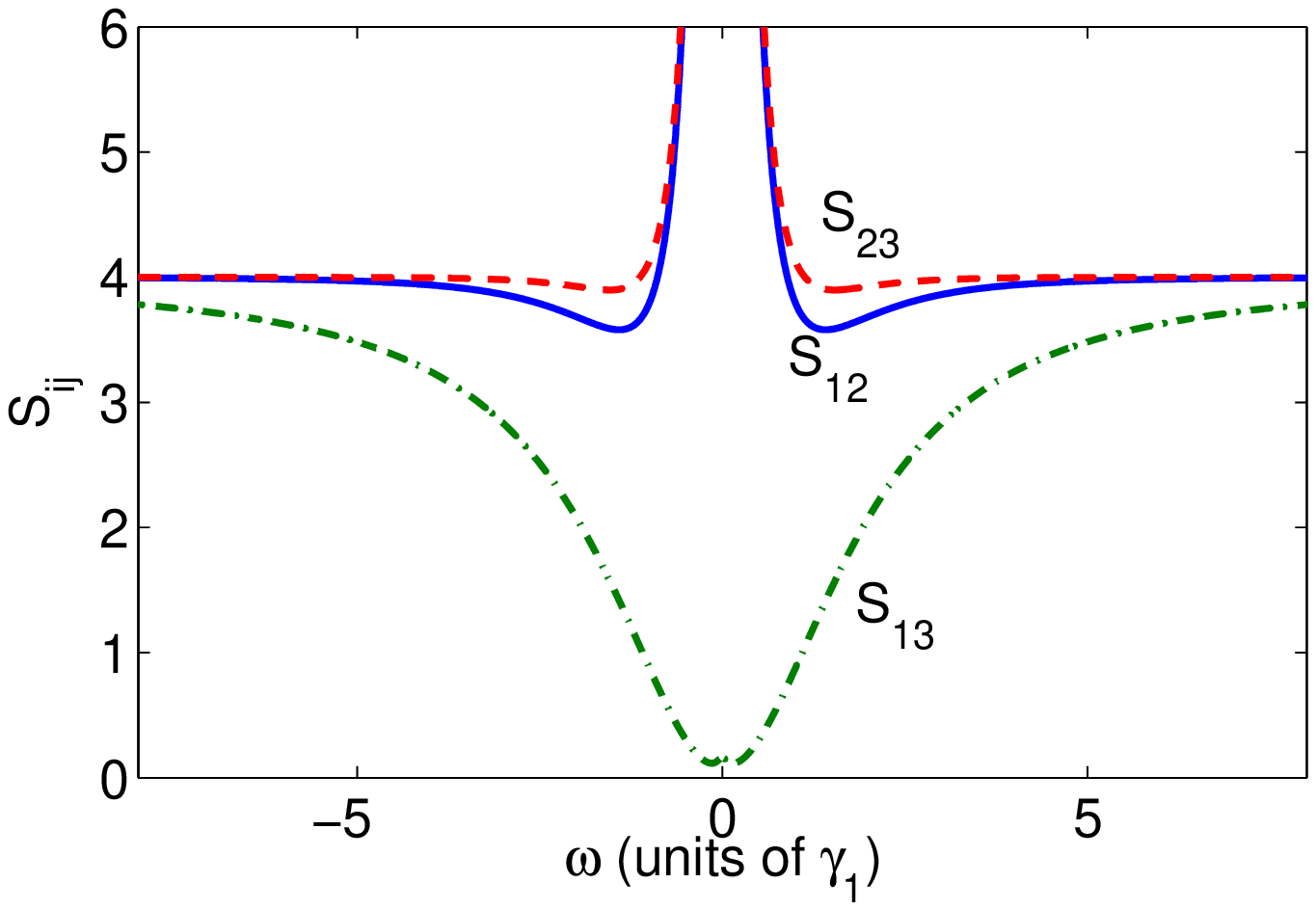}
\caption{(colour online) $S_{ij}$ criteria below threshold with $\epsilon = 0.9\epsilon_c$.}
\label{fig:experiment2}
\end{figure}

We see in Fig.~\ref{fig:experiment1} and Fig.~\ref{fig:experiment2} that, with $\epsilon=0.9\epsilon_{c}$, the violation of the inequalities has increased for two of the $S_{ijk}$, with the spectra bifurcating so that no entanglement is seen at near zero frequency. The $S_{ij}$ also show increased violation as the threshold is approached, but again not near zero frequency. The $S_{ij}$ again do not indicate full inseparability over as wide a frequency range as the $S_{ijk}$.

\begin{figure}[tbhp]
\includegraphics[width=0.75\columnwidth]{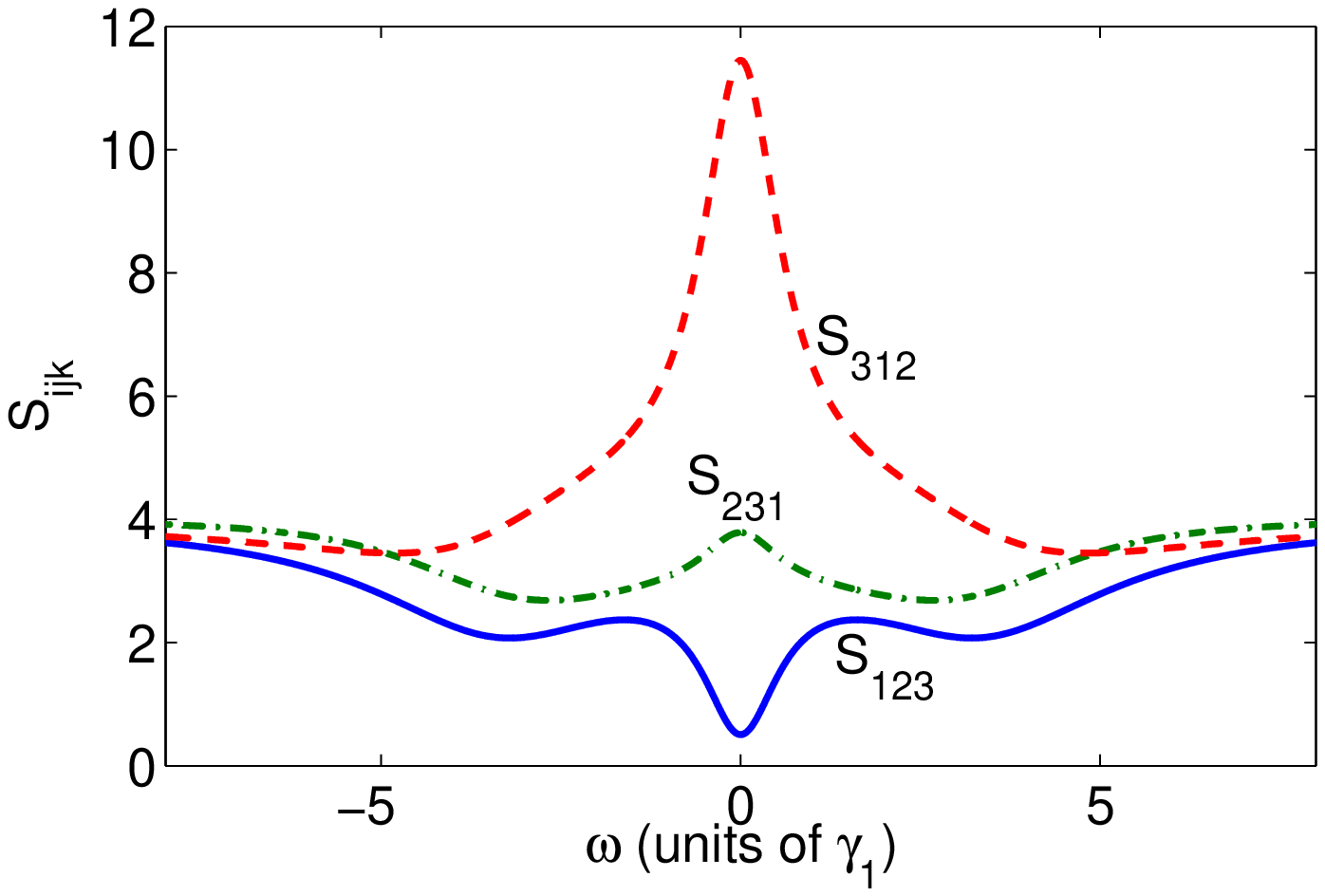}
\caption{(colour online) $S_{ijk}$ criteria for the system without threshold, with
$\epsilon = 1.5\epsilon^{o}_c$.}
\label{fig:experiment3}
\end{figure}

\begin{figure}[tbhp]
\includegraphics[width=.75\columnwidth]{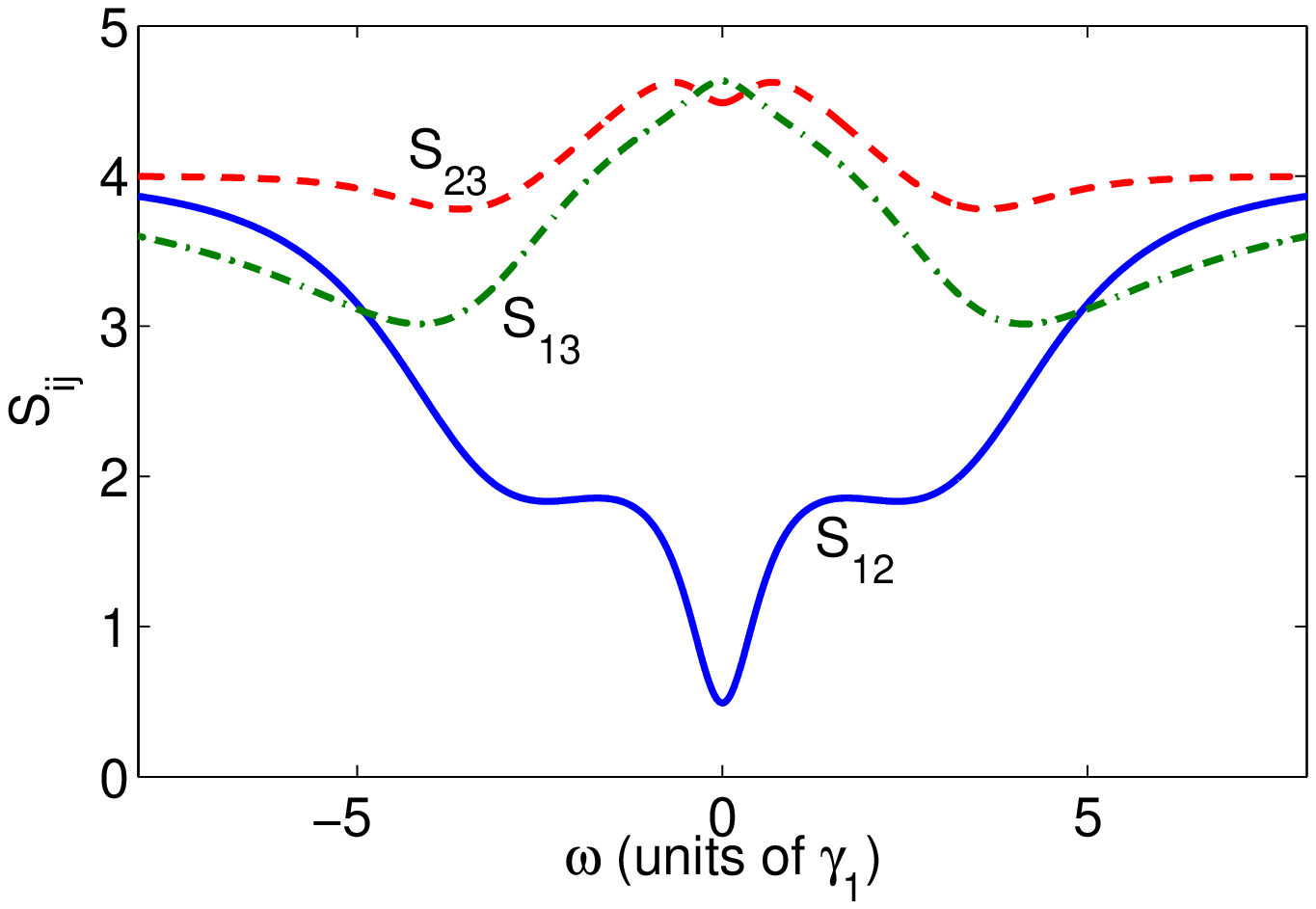}
\caption{(colour online) $S_{ij}$ criteria for the system without threshold, with
$\epsilon = 1.5\epsilon^{o}_c$.}
\label{fig:experiment4}
\end{figure}

In the region without an oscillation threshold, that is $\chi_{2}\geq \sqrt{\gamma_{2}/\gamma_{1}}\chi_{1}$, we may also apply the linearised analysis. In Fig.~\ref{fig:experiment3} and Fig.~\ref{fig:experiment4} we show the correlation functions for $\chi_2 = 2.5\chi_1$, with $\epsilon=1.5\epsilon_{c}^{o}$ and the other parameters unchanged from Fig.~\ref{fig:halfcrit1}. We see that entanglement is found in this regime and that its demonstration is less dependent on which particular correlations are measured, although $S_{312}$ and $S_{23}$ do not violate the inequality by a large amount. The main conclusion to be drawn from these results is that the $S_{ijk}$ correlations are the most appropriate to use for this system, with $S_{123}$ giving the maximum violation of the inequalities.
 
\begin{figure}[tbhp]
\includegraphics[width=.6\columnwidth]{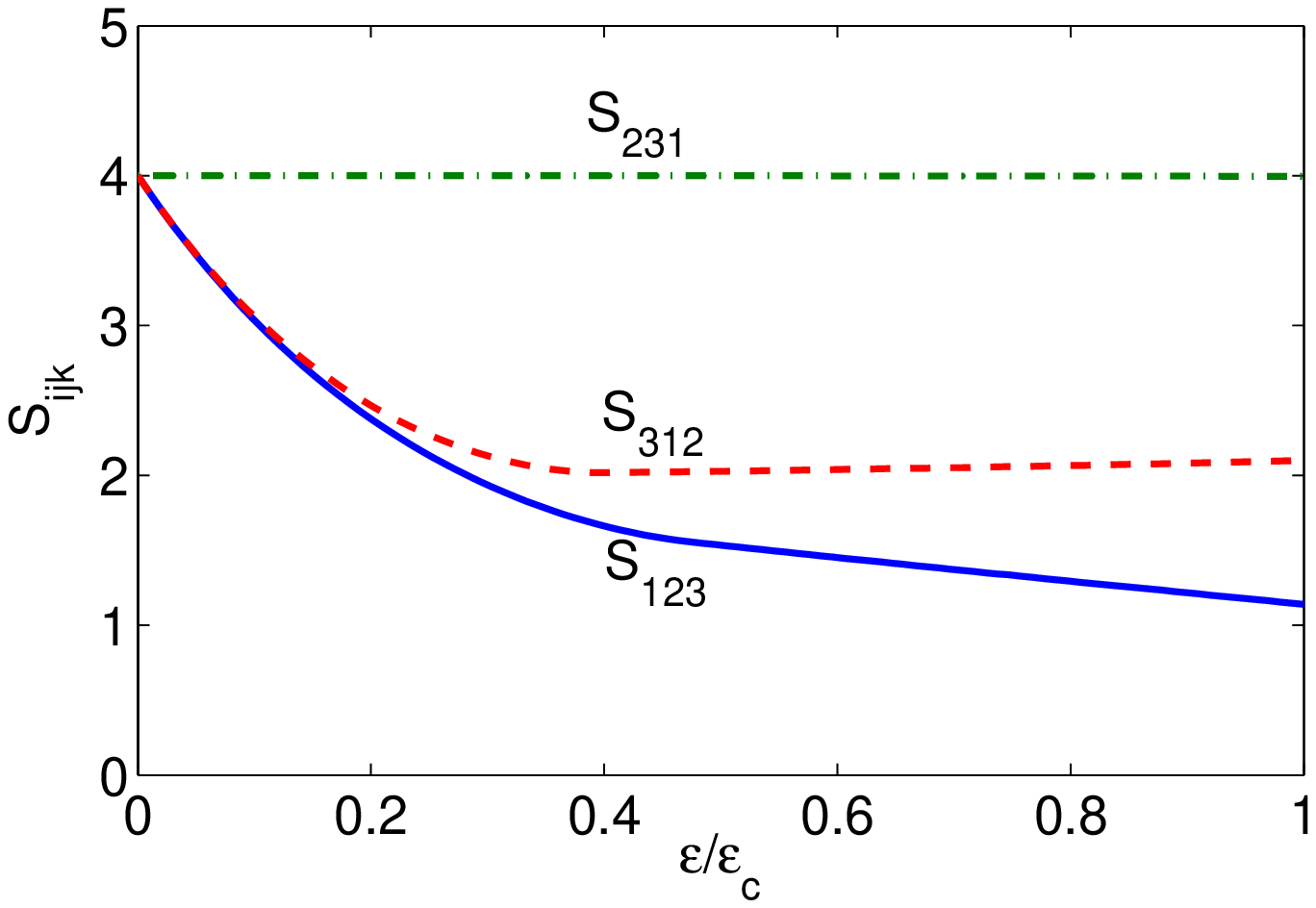}
\caption{(colour online) Minimum of the $S_{ijk}$ at any frequency between $0$ and $10\gamma_{0}$ as the pump varies up to $\epsilon_{c}$ in the region with threshold.}
\label{fig:blabla}
\end{figure}

We will now investigate the effects of changing $\epsilon$ on the correlations, concentrating on the $S_{ijk}$, as these have proven to be a more sensitive measure than the $S_{ij}$. We firstly examine the region which has a threshold, that is where $\epsilon<\epsilon_{c}$ and $\chi_{2}<\chi_{1}\sqrt{\gamma_{2}/\gamma_{1}}$. We will present these results at the frequency of maximum violation of the inequalities, with the range $0\leq\omega\leq\gamma_{0}$, rather than fixing the frequency as the other parameters are changed. In Fig.~\ref{fig:blabla} we show the minima of the three $S_{ijk}$ as the pump varies between zero and the critical value, with
$\chi_1=0.01$, $\chi_2=0.4\chi_1$, $\gamma_0 =
\gamma_1 = \gamma_3 = 1$, and $\gamma_2 = 3$. We see that in no case does $S_{231}$ violate the inequality, while the other two show clear violations, with $S_{123}$ decreasing as threshold is approached. We note here that the results in the immediate neighbourhood of the threshold are not expected to be accurate, due to the invalidity of the linearised fluctuation analysis at that point.

\begin{figure}[tbhp]
\includegraphics[width=0.75\columnwidth]{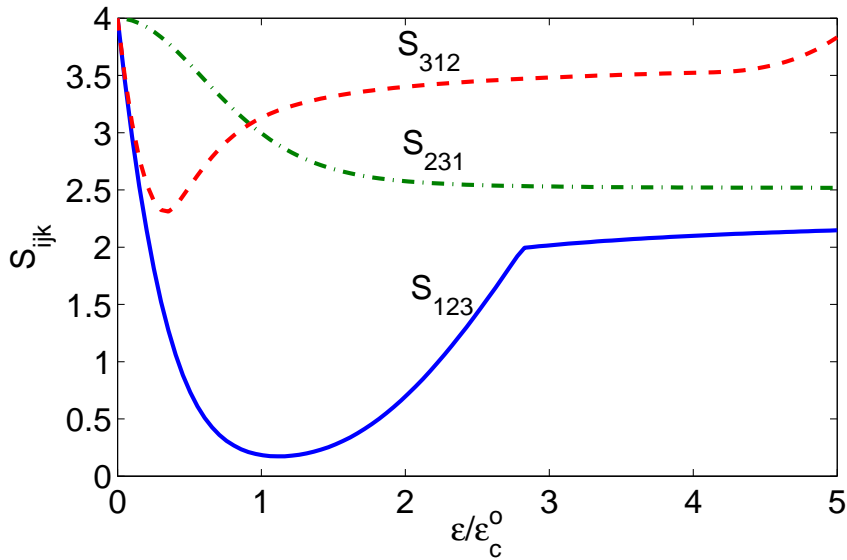}
\caption{(colour online) Tripartite entanglement criteria for the system without threshold, with $\gamma_0 = \gamma_1 = \gamma_3 =
1$, $\gamma_2 = 3$, $\chi_1 = 0.01$, and
$\chi_2 = 2\chi_1$.}
\label{fig:bombeio1}
\end{figure}

\begin{figure}[tbhp]
\includegraphics[width=.75\columnwidth]{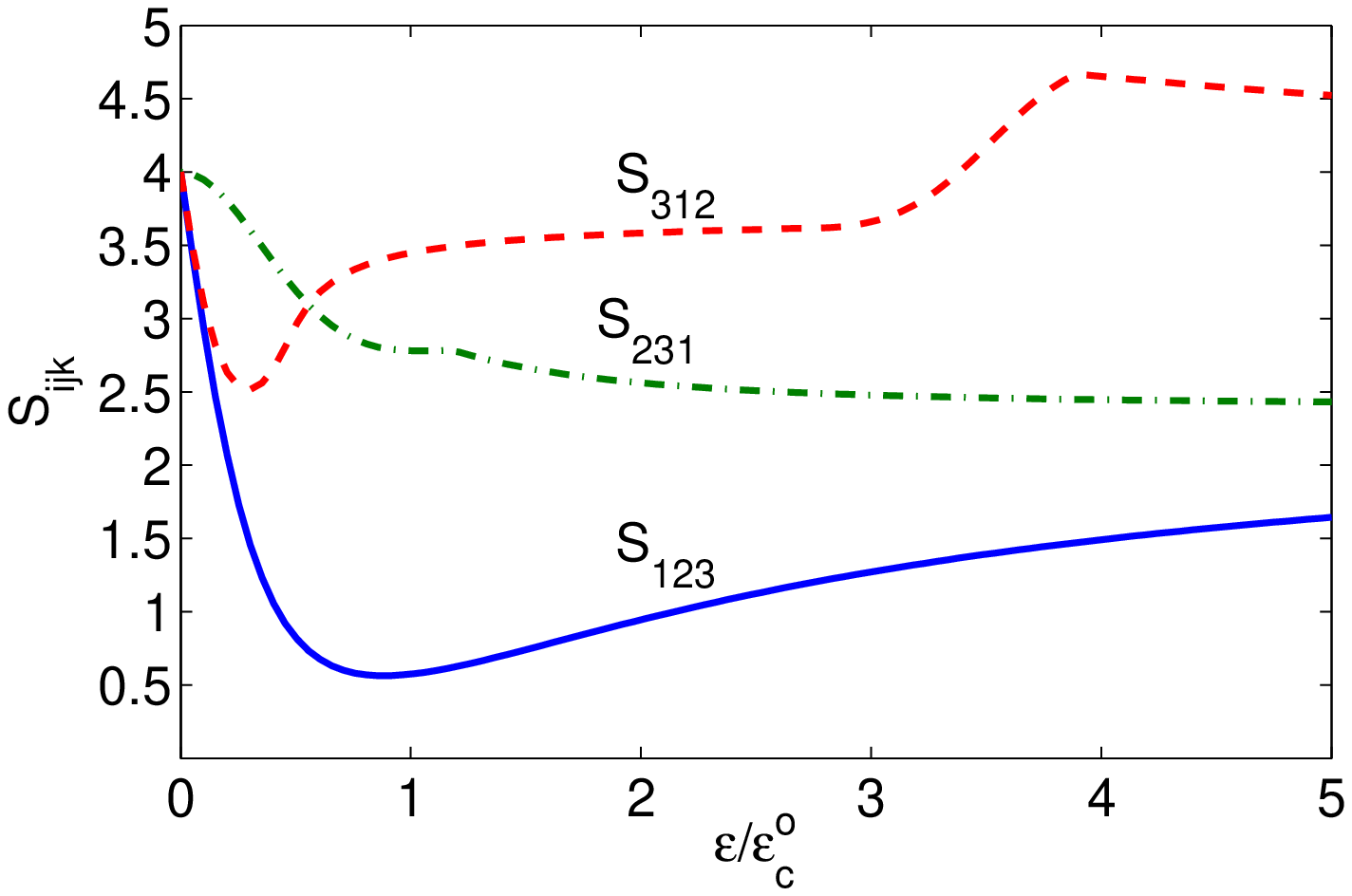}
\caption{(colour online) Tripartite entanglement criteria for the system without threshold, with $\gamma_0 = \gamma_1 = \gamma_3 =
1$, $\gamma_2 = 3$, $\chi_1 = 0.01$, and $\chi_2 = 3\chi_1$.}
\label{fig:bombeio2}
\end{figure}

\begin{figure}[tbhp]
\includegraphics[width=0.75\columnwidth]{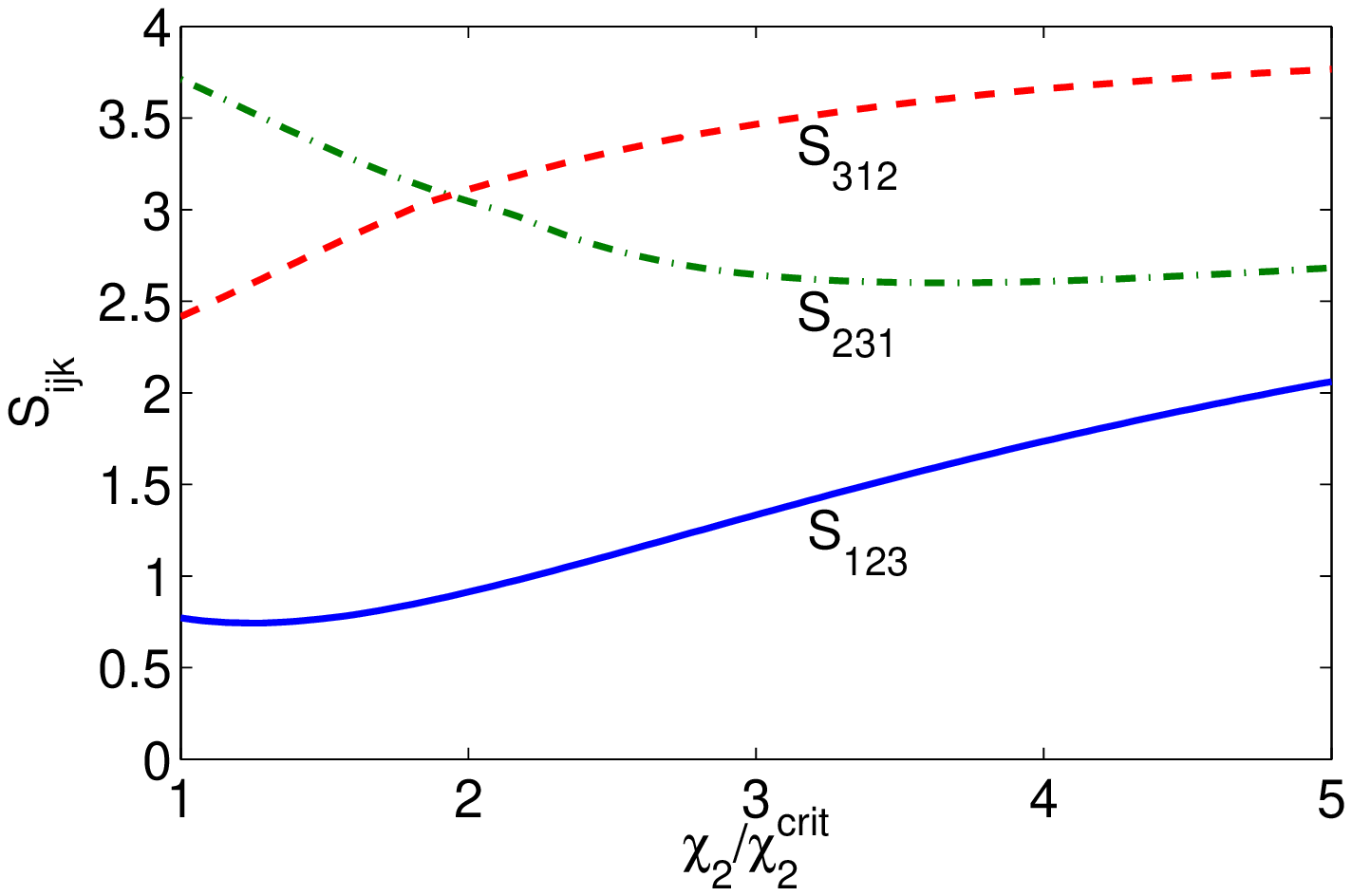}
\caption{(colour online) Tripartite entanglement criteria for the system without threshold, with $\gamma_0 = \gamma_1 = \gamma_3 =
1$, $\gamma_2 = 3$, $\chi_1 = 0.01$, and
$\epsilon = 0.5\epsilon^{o}_c$. The results are plotted as a function of $\chi_{2}^{crit}=\chi_{1}\sqrt{\gamma_{2}/\gamma_{1}}$.}
\label{fig:ee1}
\end{figure}

\begin{figure}[tbhp]
\includegraphics[width=.75\columnwidth]{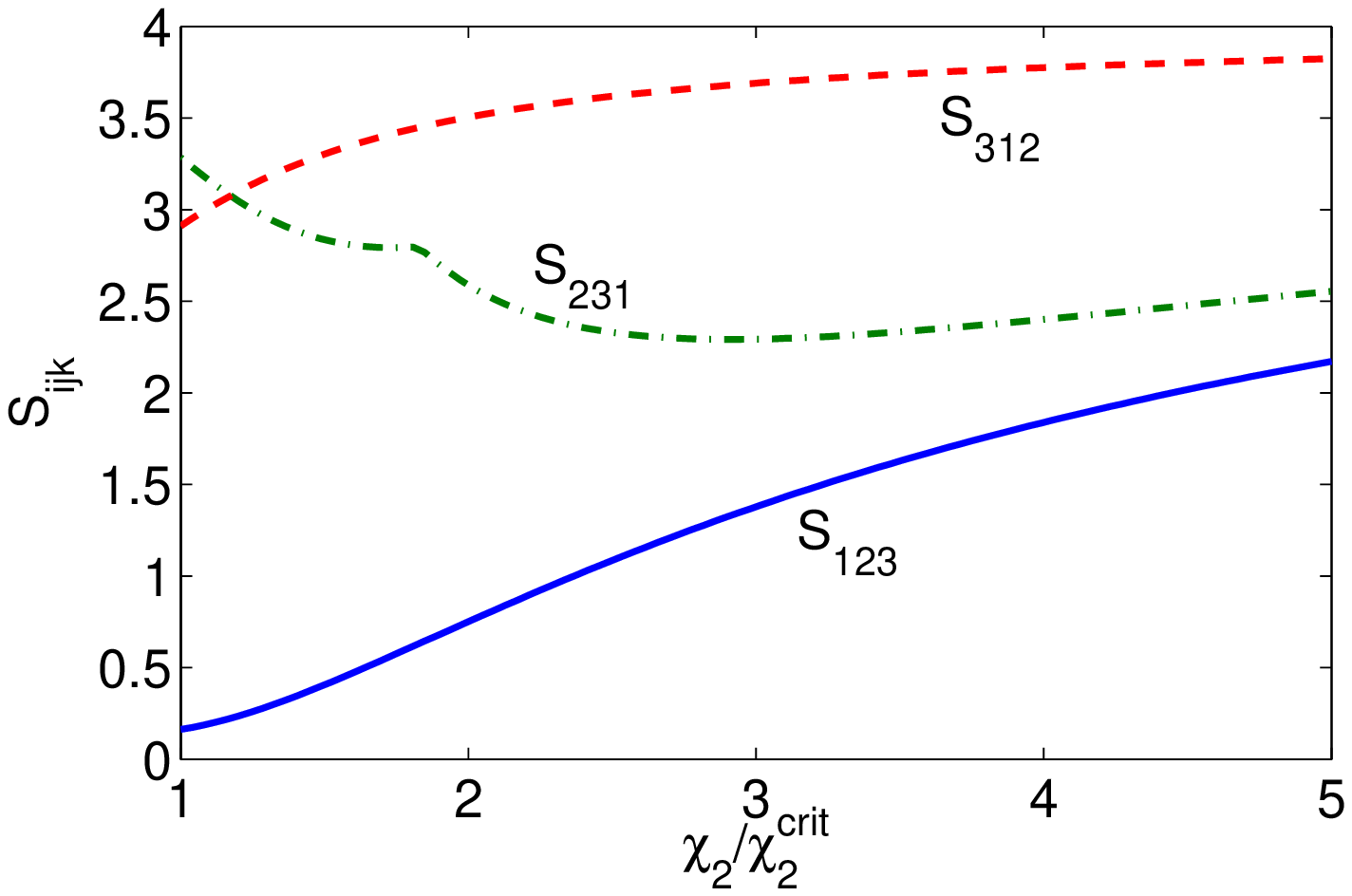}
\caption{(colour online) Tripartite entanglement criteria for the system without threshold, with $\gamma_0 = \gamma_1 = \gamma_3 =
1$, $\gamma_2 = 3$, $\chi_1 = 0.01$, and $\epsilon = 0.9\epsilon^{o}_c$. The results are plotted as a function of $\chi_{2}^{crit}=\chi_{1}\sqrt{\gamma_{2}/\gamma_{1}}$.}
\label{fig:ee2}
\end{figure}

In the parameter regime where there is no oscillation threshold, we can also investigate the effects of varying both the pumping and $\chi_{2}$.
We again set $\gamma_0 = \gamma_1 = \gamma_3 = 1$, $\gamma_2 = 3$ and $\chi_1 =
0.01$, and will allow $\omega$ to vary so as to find the maximal violations. In Fig.~\ref{fig:bombeio1} and Fig.~\ref{fig:bombeio2}, we plot the $S_{ijk}$ as a function of $\epsilon / \epsilon^{o}_c$, for values of $\chi_{2}=2\chi_{1}$ and $3\chi_{1}$. We again see that $S_{123}$ gives the maximal violations, although this does not increase monotonically with pump amplitude. 
In Fig.~\ref{fig:ee1} and Fig.~\ref{fig:ee2} we show how the correlations, again at the optimal frequencies, change as $\chi_{2}$ is increased from $\chi_{2}^{crit}$ ($=\chi_{1}\sqrt{\gamma_{2}/\gamma_{1}}$), for two different pumping amplitudes. We again see clear evidence of genuine tripartite entanglement over the range shown, with $S_{123}$ again showing the maximum violations of the inequality. We note that the violations do not increase as $\chi_{2}$ increases, but that $S_{123}$ has its minima at $\chi_{2}^{crit}$.

The results found from our semi-classical analysis show a range of different behaviours, including threshold behaviour which depends on an interplay of the two nonlinearities and could not be found in the undepleted pump treatment of Ref.~\cite{Yu}. We have been able to demonstrate that genuine tripartite entanglement is available for a wide range of parameters but have not been able to analyse the system in the above threshold regime, due to phase diffusion of the modes and the inapplicability of the linearisation procedure. In order to investigate the behaviour above threshold we will now turn to numerical stochastic integration. 

\section{Stochastic integration in the unstable regime}
\label{sec:Wigner}

In this case the numerical integration of the full positive-P equations (Eq.~\ref{eq:SDE}) presents stability problems, so that, although they were useful in deriving the correct linearised equations to calculate normally-ordered correlation functions, we will now turn to what is known as the truncated Wigner representation~\cite{Robert}.
Following the standard procedures~\cite{Crispin}, the generalised Fokker-Plank equation for the Wigner representation pseudo-probability function of the system  is found as
\begin{eqnarray}
\frac{dW}{dt} &=& \left\{-\left[\frac{\partial}{\partial\alpha_{1}}\left(\gamma_{1}\alpha_{1}-\chi_{1}\alpha_{3}^{\ast}
\beta\right) + \frac{\partial}{\partial\alpha_{1}^{\ast}}\left(\gamma_{1}\alpha_{1}^{\ast}-\chi_{1}\alpha_{3}\beta^{\ast}
\right)\right.\right.\nonumber\\
& & 
\left.\left.
+\frac{\partial}{\partial\alpha_{2}}\left(\gamma_{2}\alpha_{2}-\chi_{2}\alpha_{3}\beta\right) + 
\frac{\partial}{\partial\alpha_{2}^{\ast}}\left(\gamma_{2}\alpha_{2}^{\ast}-
\chi_{2}\alpha_{3}^{\ast}\beta^{\ast}\right)\right.\right.\nonumber\\
& &
\left.\left.
+\frac{\partial}{\partial\alpha_{3}}\left(\gamma_{3}\alpha_{3}-\chi_{1}\alpha_{1}^{\ast}\beta+
\chi_{2}\alpha_{2}\beta^{\ast}\right)
+\frac{\partial}{\partial\alpha_{3}^{\ast}}\left(\gamma_{3}\alpha_{3}^{\ast}-\chi_{1}\alpha_{1}\beta^{\ast}+
\chi_{2}\alpha_{2}^{\ast}\beta\right)\right.\right.\nonumber\\
& &
\left.\left.
+\frac{\partial}{\partial\beta}\left(\gamma_{0}\beta-\epsilon+\chi_{1}\alpha_{1}\alpha_{3}+\chi_{2}\alpha_{2}\alpha_{3}^{\ast}\right)\right.\right.\nonumber\\
& &
\left.\left.
+\frac{\partial}{\partial\beta^{\ast}}\left(\gamma_{0}\beta^{\ast}-\epsilon^{\ast}+\chi_{1}\alpha_{1}^{\ast}\alpha_{3}^{\ast}+\chi_{2}\alpha_{2}^{\ast}\alpha_{3}\right)\right]\right.\nonumber\\
& & 
\left.
+\frac{1}{2}\left[\frac{\partial^{2}}{\partial\alpha_{1}\partial\alpha_{1}^{\ast}}(2\gamma_{1})+\frac{\partial^{2}}{\partial\alpha_{2}\partial\alpha_{2}^{\ast}}(2\gamma_{2})+\frac{\partial^{2}}{\partial\alpha_{3}\partial\alpha_{3}^{\ast}}(2\gamma_{3})+\frac{\partial^{2}}{\partial\beta\partial\beta^{\ast}}(2\gamma_{0})\right]\right.
\nonumber\\
& &
\left.
-\frac{1}{8}\left[\frac{\partial^{3}}{\partial\alpha_{2}\partial\alpha_{3}^{\ast}\partial\beta^{\ast}}(2\chi_{2})+\frac{\partial^{3}}{\partial\alpha_{2}^{\ast}\partial\alpha_{3}\partial\beta}(2\chi_{2})\right.\right.
\nonumber\\
& &
\left.\left.
-\frac{\partial^{3}}{\partial\alpha_{1}\partial\alpha_{3}\partial\beta^{\ast}}(2\chi_{1})-
\frac{\partial^{3}}{\partial\alpha_{1}^{\ast}\partial\alpha_{3}^{\ast}\partial\beta}(2\chi_{1})
\right]
\right\}W.
\label{eq:Wfokkerplanck}
\end{eqnarray}

We immediately see that the above equation contains third-order derivatives so that it cannot be mapped onto a set of stochastic differential equations. Although methods have been developed to map these type of generalised Fokker-Planck equations onto stochastic difference equations in a doubled phase space~\cite{BWO}, the integration of these can present more stability problems than the positive-P representation, so we will not pursue this approach here. Hence we neglect the third-order derivatives to allow a mapping onto the set of stochastic equations in the truncated Wigner representation,
\begin{eqnarray}
\frac{d\alpha_1}{dt} &=& -\gamma_1\alpha_1 + \chi_1\alpha_3^{\ast}\beta +
\sqrt{\frac{\gamma_1}{2}}(\eta_1 +
i\eta_2),\nonumber\\
\frac{d\alpha_2}{dt} &=& -\gamma_2\alpha_2 + \chi_2\alpha_3\beta +
\sqrt{\frac{\gamma_2}{2}}(\eta_3 +
i\eta_4),\nonumber\\
\frac{d\alpha_3}{dt} &=& -\gamma_3\alpha_3 + \chi_1\alpha_1^{\ast}\beta - \chi_2\alpha_2\beta^{\ast} +
\sqrt{\frac{\gamma_3}{2}}(\eta_5 +
i\eta_6),\nonumber\\
\frac{d\beta}{dt} &=& \epsilon - \gamma_0\beta - \chi_1\alpha_1\alpha_3 - \chi_2\alpha_2\alpha_3^{\ast} +
\sqrt{\frac{\gamma_0}{2}}(\eta_7 +
i\eta_8),
\label{eq:stochastic2}
\end{eqnarray}
where the $\eta_{j}$ are Gaussian random noises as defined by Eq.~\ref{eq:ruido}.
As well as not containing multiplicative noise terms, another important difference from the positive-P equations is that the initial conditions on each stochastic trajectory must be drawn from the appropriate Wigner distribution for the desired quantum state of the mode. We will be beginning our trajectories with vacuum inside the cavity, so that, for example, we choose $\alpha_{j}^{n}(0)=(\xi_{1}^{n}+i\xi_{2}^{n})/2$ on the $n$th trajectory (and similarly for $\beta(0)$), where the $\xi$ are normal Gaussian random numbers with zero mean. The Wigner representation naturally calculates symmetrically-ordered operator averages, so that care must be taken with any necessary reordering to give predictions for observables. We also note here that, while cases have been found where the truncated Wigner representation can give inaccurate results~\cite{punheteiro,arabe}, we expect it to be accurate here because all four modes are macroscopically occupied in the region that we are using it to investigate. In our stochastic integration we have set $\gamma_0 = \gamma_1 = \gamma_3 = 1$,
$\gamma_2 = 3\gamma_1$, $\chi_1 = 0.01\gamma_1$, $\chi_2 = 0.4\chi_1$ and $\epsilon=1.5\epsilon_{c}$. An indication of the accuracy is that it gives predictions for the intracavity field intensities that are consistent with the analytical values given above, as shown in the table.

\begin{center}
\begin{tabular}{|c||c|c|}
\hline
 & Analytic & Wigner \\
\hline
$|\beta|^{2}$ & $1.056\times 10^{4}$ & $1.056\times 10^{4}$\\
$|\alpha_{1}|^{2}$ & $5.0143\times 10^{3}$ & $5.0141\times 10^{3}$\\
$|\alpha_{2}|^{2}$ & $89.1424$ & $89.0909$\\
$|\alpha_{3}|^{2}$ & $4.7468\times 10^{3}$ & $4.7471\times 10^{3}$\\
\hline
\end{tabular}
\end{center}

\begin{figure}[tbhp]
\includegraphics[width=.75\columnwidth]{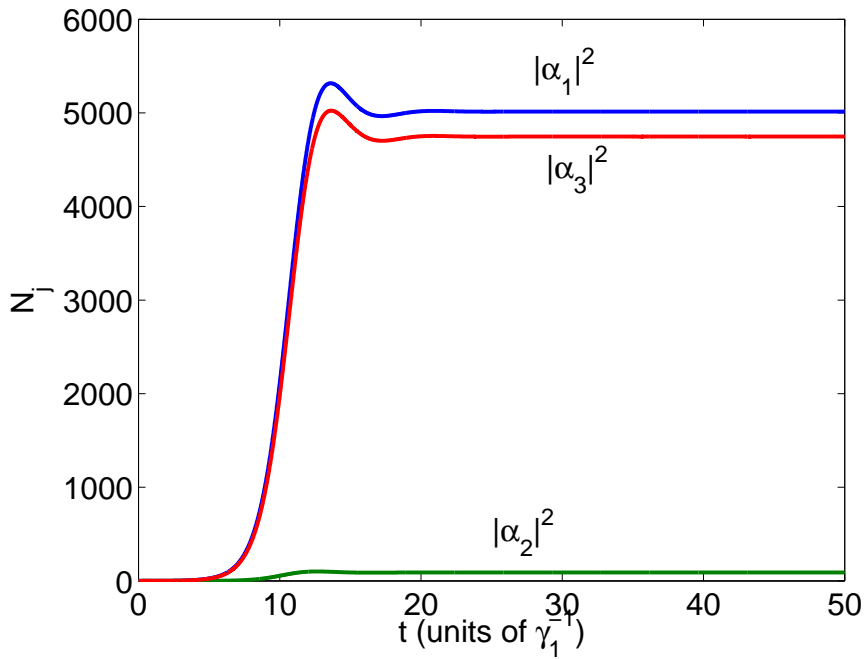}
\caption{(colour online) Intensity of the output modes $|\overline{\alpha_i|^2}$, obtained by
stochastic integration of the truncated Wigner equations averaged over $2.65\times 10^{5}$ trajectories.
Note that the horizontal axis is now time scaled in units of $\gamma_{1}^{-1}$.}
\label{fig:wig1}
\end{figure}

\begin{figure}[tbhp] 
\includegraphics[width=.75\columnwidth]{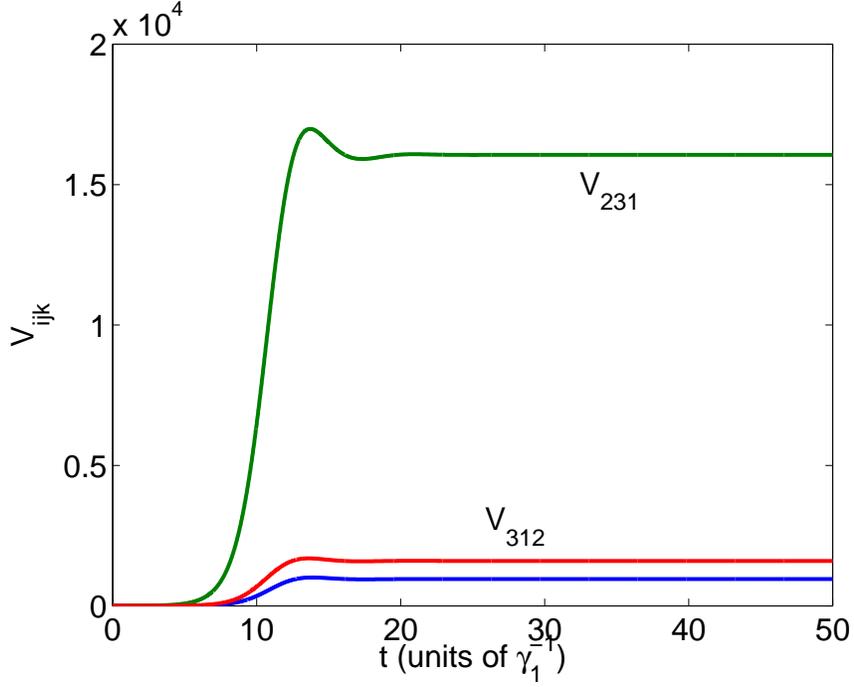}
\caption{(colour online) The $V_{ijk}$ for $\epsilon = 1.5\epsilon_c$, obtained by
stochastic integration of the truncated Wigner equations averaged over $2.65\times 10^{5}$ trajectories.
Note that the horizontal axis is now time scaled in units of $\gamma_{1}^{-1}$.}
\label{fig:wig1a}
\end{figure}

\begin{figure}[tbhp]
\includegraphics[width=.6\columnwidth]{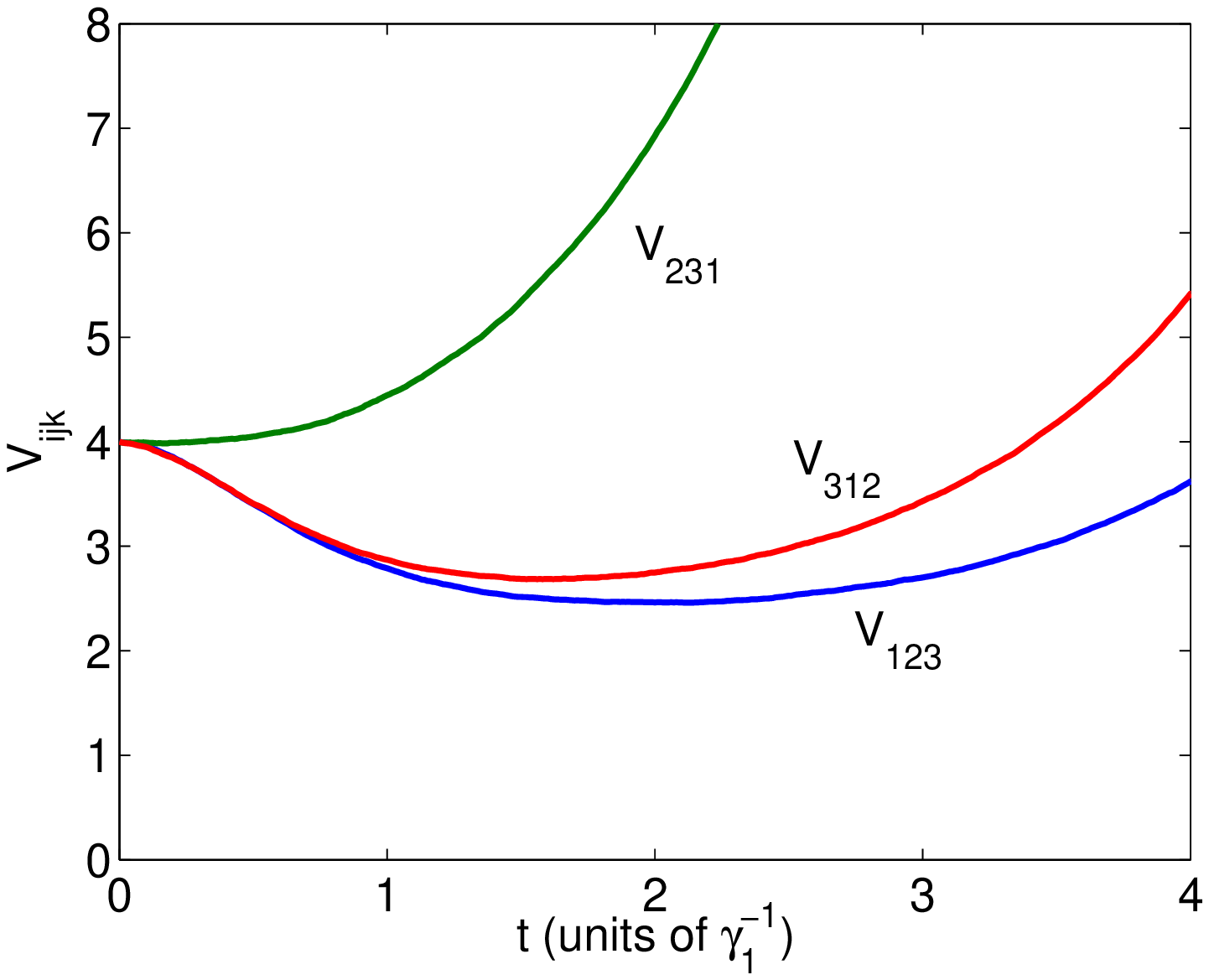}
\caption{(colour online) Transient behaviour of the $V_{ijk}$ with $\epsilon = 1.5\epsilon_c$. As seen in Fig.~\ref{fig:wig1}, the violations of the inequalities persist for only a short time.}
\label{fig:wig2}
\end{figure}

In Fig.~\ref{fig:wig1} and Fig.~\ref{fig:wig1a} we show the intensity of the $\alpha$ modes inside the cavity, demonstrating both that we have reached the steady-state regime and that the modes are macroscopically occupied. Integration for values of $\epsilon$ close to $\epsilon_{c}$ typically took much longer to reach the steady-state, due to critical slowing down, a well-known phenomenon which occurs in the vicinity of phase transitions. Due to the phase diffusion predicted in the analytical solutions, the averages of the $\alpha_{j}$ are essentially zero and no entanglement is registered by the $V_{ijk}$ correlations in the steady-state, with these all being far from violating the inequalities. As shown in Fig.~\ref{fig:wig2}, there is some violation in the initial transient regime before the pump mode within the cavity builds up to its threshold value and the system begins to oscillate. This transient feature is unlikely to be of any practical use, as it exists for only a few cavity lifetimes and the fields are no more intense than in the below threshold regime, where genuine tripartite entanglement is readily seen in the steady-state regime. We note here that the three-mode EPR correlations~\cite{EPR3} give much smaller values in the steady-state, but still do not violate the inequalities. It is possible that a small injected signal at $\omega_{2}$ could serve to lock the phases and enable entanglement to be observed, but investigation of this is outside the scope of the present work.

\section{Conclusion}

We have analysed the system of singly-pumped intracavity coupled downconversion and sum-frequency generation with a quantised pump field. Unlike previous analyses, this enabled us to define the threshold properties of the system and analyse the dynamics with all modes oscillating macroscopically inside the cavity. One of the features we have found is that for some values of the
experimental parameters, the threshold value of the pump field diverges so that, however strongly the cavity is pumped, the system will not oscillate. We have found that genuine tripartite entanglement is available in both the regions below and without threshold, but that, as in other systems with asymmetric Hamiltonians, not all measurable correlations will detect the violation of the entanglement inequalities. Above threshold the converted modes undergo phase diffusion, which prevents the detection of entanglement based on quadrature measurements except in the early transient regime. This signifies that the system is not a good candidate for the production of bright entangled output beams, unless a method can be found to overcome the problem of phase diffusion. However, it is still useful for the production of genuine polychromatic tripartite entanglement in all except the above threshold regime.

\section*{Acknowledgments}

This work was supported by the Australian Research Council and the Queensland state government. C. Pennarun was a guest of the University of Queensland with financial support from the French government programme Explo'ra Sup. 


\end{document}